# THE COBRAS/SAMBA CMB PROJECT

François R. Bouchet[1], Richard Gispert[2], and Jean-Lou Puget[2]
[1] *Institut d'Astrophysique de Paris, CNRS, Paris, France.*
[2] *Institut d'Astrophysique Spatiale, Orsay, France.*

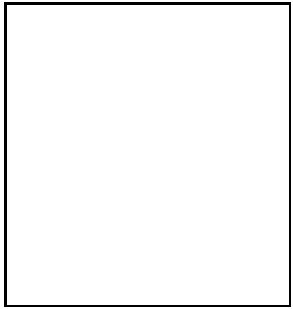

**Abstract**

COBRAS/SAMBA is a second generation satelitte dedicated to mapping at high resolution and sensitivity the anisotropies of the Cosmic Microwave Background (CMB). This mission is in the assessment study phase (A) at ESA, with a decision expected mid 1996, for a launch around 2003.

## 1 Small scale anisotropies of the Cosmic Microwave Background

A central problem in cosmology is the building and testing of a full and detailed theory for the formation of (large-scale) structures in the Universe. It is widely believed that the observed structures today grew by gravitational instability out of very small density perturbations. Such perturbations should have left imprints as small temperature anisotropies in the cosmic microwave background (CMB) radiation.

The COBE satellite has opened up a new era by reporting the first detection of such temperature anisotropies. However, COBE was limited by poor sensitivity, and restricted to angular scales greater than 7 degrees, much more extended than the precursors of any of the structures observed in the Universe today. As a result, these measurements are compatible with a wide range of cosmological theories. On the other hand, high precision measurements at the degree scale will be extremely discriminating, allowing strong constraints on large scale structure formation theories, the ionization history, the cosmological parameters (such as the Hubble and cosmological constants, or the type and amount of dark matter), and very early Universe particle theories. These goals could be met by producing nearly full-sky maps of the background anisotropies with a sensitivity of $\frac{\Delta T}{T} \sim 10^{-6}$ on scales from 10 arc minutes to tens of degrees.

More precisely, the <u>statistics of the background anisotropies</u>, if Gaussian, would favor inflationary models, whereas non-Gaussian fluctuations would rather favor models in which irregularities are generated by topological defects such as strings, monopoles and textures. The shape of the primordial fluctuation spectrum, according to most theories of the origin of the fluctuations in the Universe corresponds to potential fluctuations $\delta\phi$ which should be independent of the scale of the irregularities ($\lambda$). Even a few percent deviation from this prediction would have extremely important consequences for the inflationary models. These models also predict a specific ratio of tensor (generated by gravitational waves) vs. scalar anisotropies depending notably on the scale of the fluctuations. The best way to learn about the reionization history of the universe would be through the small-scale anisotropies which may be partly erased if the inter-galactic medium was re-ionized at high redshift. These small scale anisotropies depend also on the baryon density (see fig. 1.a), the nature of the dark matter, and the geome--try of the Universe (and of course on the initial spectrum of irregularities). Observations of the structure of the CMB on scales down to $10'$ will resolve structures comparable in scale to clusters of galaxies ($\sim 10 h^{-1} \mathrm{Mpc}^1$) and so allow a much more direct link between observations of galaxy clustering, galaxy peculiar velocities and temperature anisotropies. Additionally, the detection of at least 1000 rich clusters[2] by the Sunyaev-Zeldovich effect[4], in combination with X-ray observations would constrain the evolution of rich clusters of galaxies and provide another handle on $H_0$. Finally, the Doppler effect on the same clusters might provide a unique measurement of their peculiar velocity dispersion.

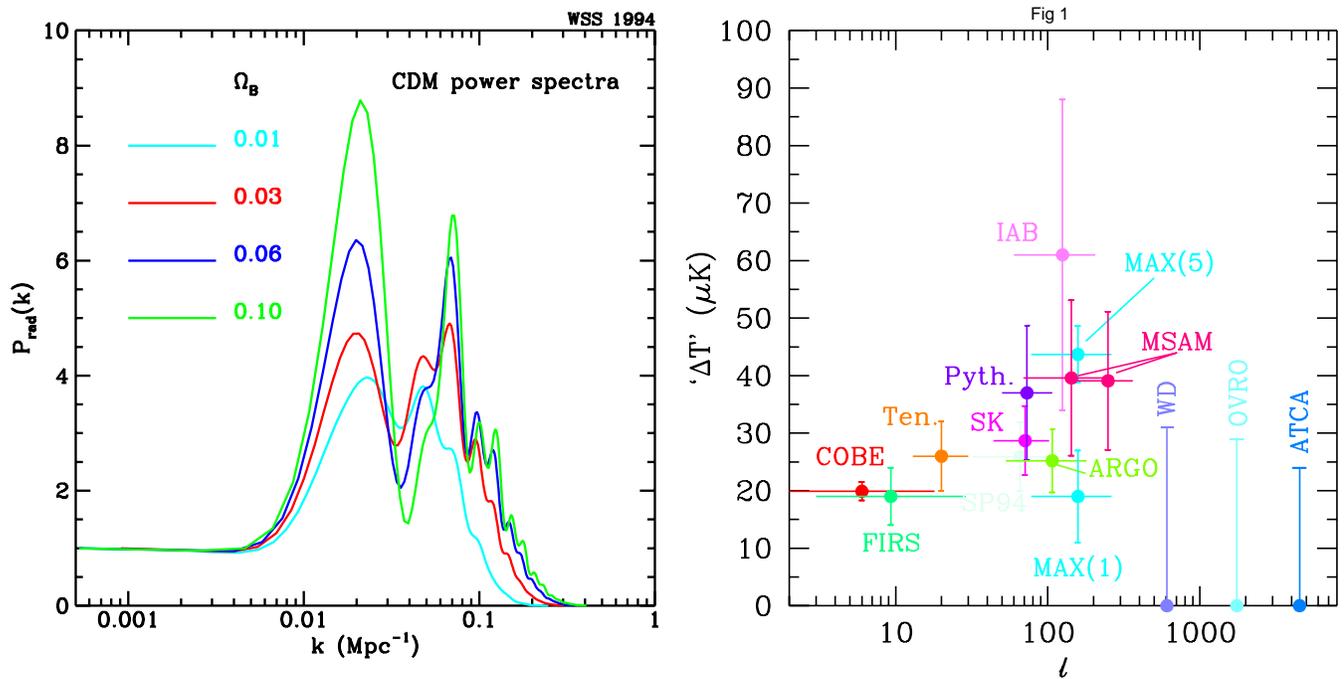

Figure 1: (a) Variation with the baryonic content of the Universe of the power spectrum for a CDM model(b) Currrent state of the CMB measurements. Both plots are courtesy of the Berkeley CMB group.

---

[1] $h$ is Hubble's constant in units of 100 km s$^{-1}$Mpc$^{-1}$.
[2] assuming a photon noise limited space experiment.

# 2 Model of the Galactic Foregrounds

The final accuracy of the measurements of the cosmological anisotropies will rely on the ability to separate these fluctuations from those coming from the galactic background and from unresolved extra galactic sources. It is thus crucial to know the expected contribution of these foregrounds, in order to predict the expected fraction of the sky for which no "cleaning" would be necessary. Here we describe our model for the galactic emission to do this.

## 2.1 Dust emission

The spectrum of the dust emission in the wavelength range of interest has been measured by the FIRAS instrument aboard COBE with a 7 degree beam (Wright et al. 1991, Reach et al. 1995). Several ballon experiments have detected the dust emission at high galactic latitude in a few photometric bands with angular resolution from 30 arc minute to 1 degree (Fischer et al. 1995, Page et al. 1990, De Bernardis 1990, Andreani 1990, Meinhold and Lubin 1991, Cheng et al. 1994...). The average spectrum at high latitude exhibits a cold component and can be modeled[7] with a dust at 21.3 K, with an emissivity $\nu^\alpha$ of spectral index $\alpha = 1.28$. We used the DIRBE data at $240\mu$ as a spatial template, and the average spectrum above to predict the long wavelenth behaviour of the dust.

This is a worst case because it assumes that the "cold component" is all interstellar and is present wherever dust emission is detected at $240\mu$. Indeed, the decomposition in two components with two temperatures (and $\alpha = 2$) in the paper by Reach et al. (1995) shows that the properties of this cold component are variable; the ratio of the optical depth of the cold to warm component increases at high latitudes. It is likely that part of this cold component is in fact extragalactic. In that case it is also a foreground to the CMB, but with quite different properties in term of inhomogeneities. The interstellar cold component is then concentrated in the higher column density lines of sight of the ISM, the more diffuse phase having an emissivity law with $\alpha = 2$. This reduces the contribution at long wavelength by a significant factor (2.7 for 1mm, 5.8 at 3mm) as compared to our pessimistic modelisation.

We checked that the power spectrum of the DIRBE fluctuations is decreasing like the $3^{rd}$ power of the spatial frequency, $\ell$, in agreement with the determination of Gautier et al. (198?) who used instead the $100\mu$ IRAS data down to a resolution of 4 arc minutes.

## 2.2 Free-free emission emission

The ionized gas in the interstellar medium can be traced in several ways, e.g. through its $H_\alpha$ emission (Reynolds 1992), the NII fine structure line at $205\mu$, the FIRAS data, or pulsar dispersion measures.

The NII measurements lead to a somewhat higher emission measure than the $H_\alpha$, the latter observations being biased toward low column density regions. Taking this into account, Bennet et al. (1992) deduce an average brightness of the free-free emission at the galactic pole of 7 $\mu$K at 53 GHz. The power spectrum of the very limited available data from Reynolds appears very similar to the $\ell^{-3}$ one of the dust. Furthermore, this $H_\alpha$ data shows a part correlated with the HI clouds traced by their 21 cm emission line.

We make the hypothesis that all the ionised Hydrogen of the ISM is associated with dust, as suggested by the observations of individual ionised regions. We can thus use the DIRBE 240 micron map as a spatial template for the dust *and* the free-free emissions correlated with HI. In addition, we model the part uncorrelated with HI by using the same template but north/south inverted; this second, uncorrelatred, component accounts for 5% of the total gaz and 50% of the free-free emission. The spectrum of the free-free emission is modeled as $I_\nu \propto \nu^{-0.16}$, and normalised to give 7 $\mu$K at 53 GHz for a dust emission of $\simeq 2$ MJy/sr at $240\mu$.

---

[7]This corresponds to the region B4 with $37 < b < 53$ and $-53 < b < -37$ in the PhD thesis of Barnes. It is also compatible with the analysis of Reach et al. 95, who find $T = 22.2$, $\alpha = 1.12$ between 30 and 60 degrees, and $T = 20.7$, $\alpha = 1.36$ between -30 and -60 degrees.

## 2.3 Synchrotron emission

We use a spectral index of $\alpha = -0.7$ to extrapolate at higher frequency the spatial template provided by the full sky map at 408 MHz of Haslam et al. (1970). We found that the power spectrum of that map at high latitude is roughly $\propto \ell^{-1.5}$. Since it is somewhat shallower than -2, its integral in a restricted range of angular scales is dominated by the small scale limit, contrary to the case of dust and free-free distributions.

## 2.4 Characterisation of the fluctuations

By using the model outlined above, one can generate at any desired frequency full sky maps with (for instance) a 1 degree FWHM resolution for the dust, free-free, and synchrotron emissions, and for their sum. We found convenient to express all the emissions in equivalent thermodynamical temperature fluctuations.

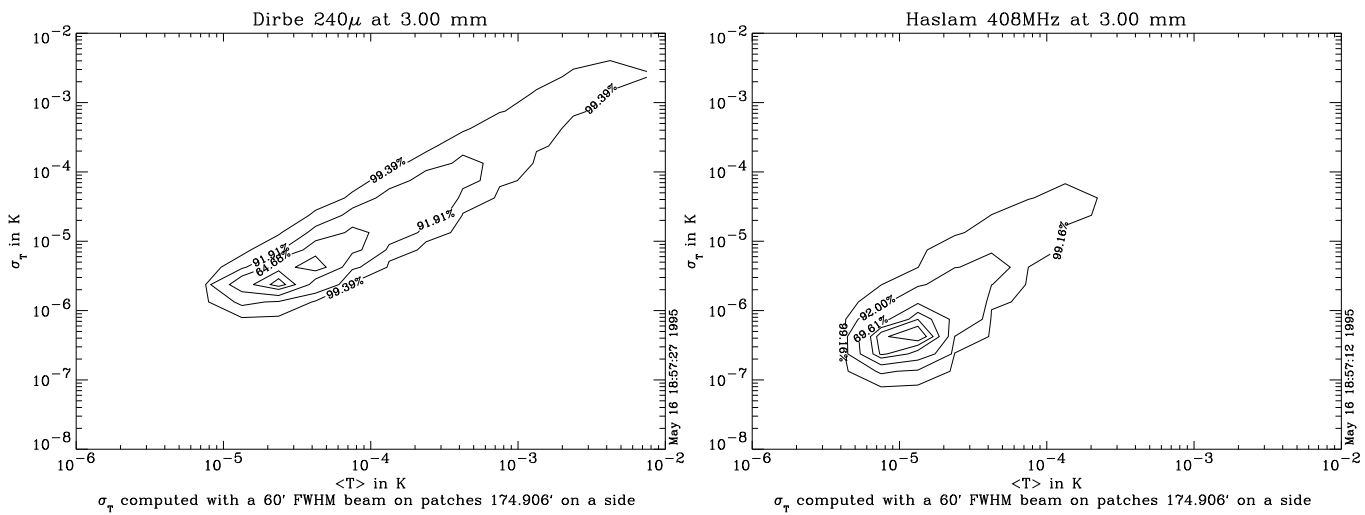

Figure 2: Temperatures distribution at 3mm for (a) "Dust+Free-Free" (b) "Synchrotron".

In order to obtain *local* estimates of the level of fluctuations at intermediate angular scales of the various components, we compute the *rms* fluctuation of the maps over square patches with 3 degrees on a side (i.e. containing about 10 beams of 1 degree FWHM). Since this amounts to restrict the contribution to the variance to a range $\ell$ ($\delta\ell$, comparable to $\ell$), we are in effect measuring the local variations of the normalisation of the angular power-spectrum. As shown in figure 2, at 3mm, 70% of the patches for the synchrotron have a mean temperature less than $2 \times 10^{-5}$, and fluctuations $\sigma_T$ about 20 times smaller. For the dust, 65% of the patches have a mean temperature less than $2 \times 10^{-4}$, and fluctuations $\sigma_T$ about 10 times smaller. This figures provide direct graphical evidence that a large fraction of the sky will have low levels of pollution.

The figure 3 gives the fraction of the sky for which the *rms* fluctuations of the foregrounds (integrated over a restricted range in $\ell$) and of their total contribution are less than a given value. In the 1 to 3 degree resolution range, the *rms* total for the best half of the sky is $1. \times 10^{-5}$ at 3mm and $3 \times 10^{-5}$ at 1.3mm. For a CDM case, the coresponding CMB fluctuations would be of order $3. \times 10^{-5}$. Note that while at 1.3 mm the fluctuations come only from the dust, at 3mm the free-free part is substancial.

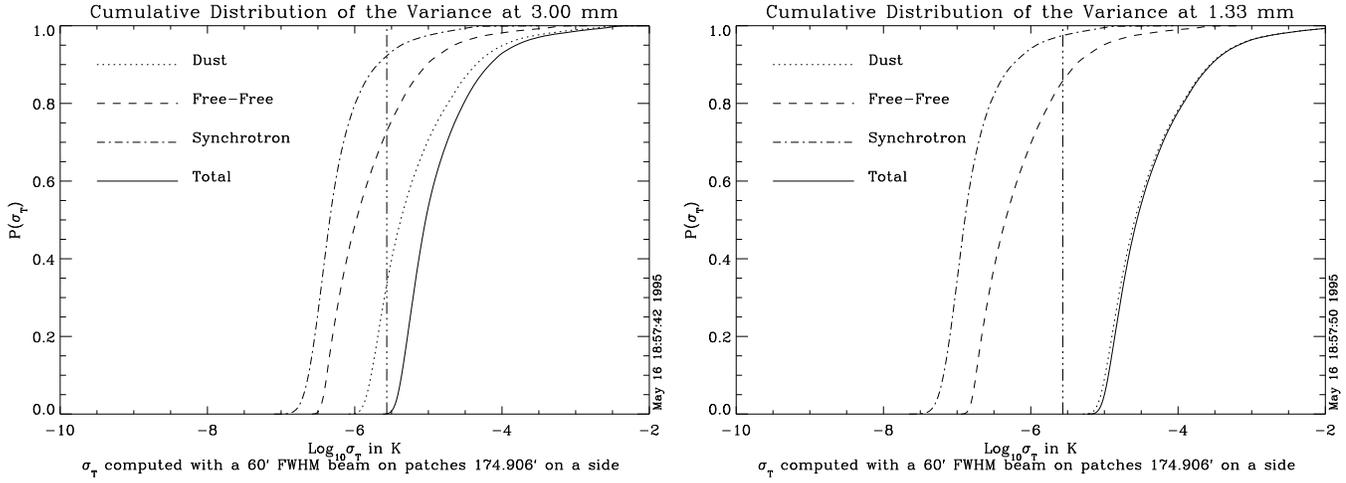

Figure 3: Foreground fluctuations in 3 degree patches for a 1 degree FWHM gaussian beam (a) at 3mm (b) at 1mm.

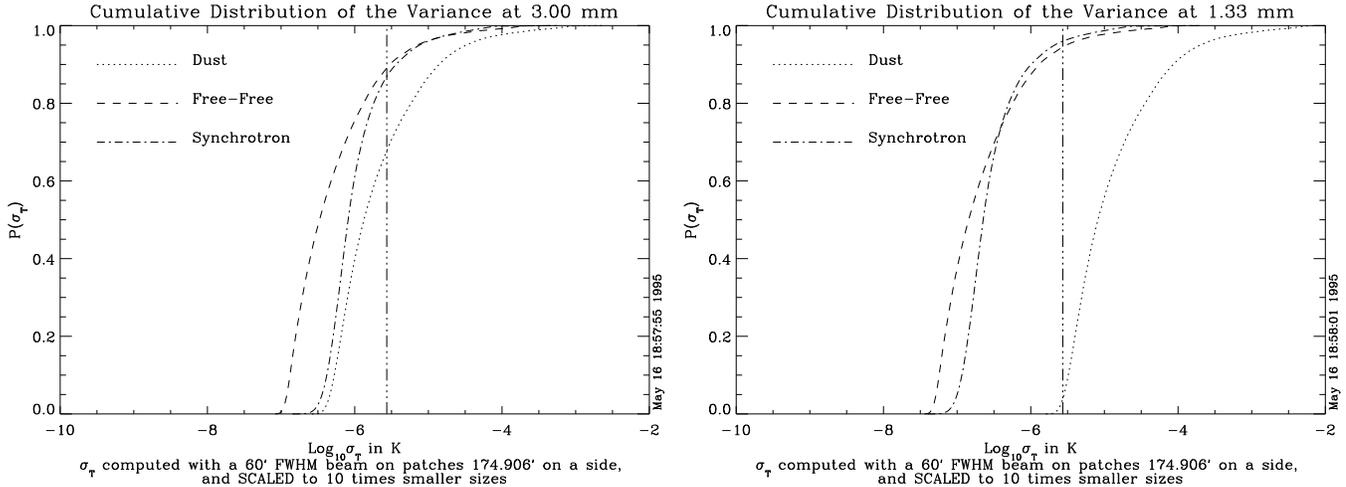

Figure 4: Foreground fluctuations scaled to 18 arc minutes patches. with a 6 arc minutes FWHM beam (a) at 3mm (b) at 1mm.

We can furthermore use the knowledge of the slope of the power spectra of the components to estimate the results of such measurements on smaller angular scales. Indeed, measurements with a 6 arc minute beam over 18 arc minute patches should be $10^{-1/2}$ smaller (dust and free-free) and $10^{1/4}$ larger (synchrotron) than the previous ones on 10 times larger scales. The figure 4.b shows such plots scaled to 10 time smaller sizes, i.e. patches of 18 arc minute for a 6 arc minute FWHM beam. In that case, the foreground gets smoother (half of the sky is with *rms* fluctuations smaller than $10^{-6}$ at 3mm, and $3. \times 10^{-6}$ at 1.3mm). CMB fluctuations in the CDM case would still be of order $3. \times 10^{-5}$. At 3mm, the contribution to the fluctuations from synchrotron and dust emission are comparable.

## 2.5 Consequences

This shows that the scientifically important small scales (structure of the "Doppler" peaks past the first one[8]) can be reached more easily at high frequency (for a given telescope size). Indeed, by working at low frequency, the foreground is dominated by the free-free that cannot be taken out because it is measurable only at longer wavelengths where the resolution is not as good. On the contrary, the dust emission gets smoother at small scales and it can be measured with a better angular resolution at higher frequency. A correction with only a 20% accuracy, which is relatively easy brings the dust contribution below the free free one, even at 3 mm.

About the extragalactic background due to early galaxies: it is possible (and even likely!) that the fluctuations due to distant galaxies dominate over those coming from the galactic dust distribution in the low cirrus regions. It is important, both for the measurement of the small scales primordial fluctuations and because of its own cosmological interest, to extract properly this component. The only way to separate it is to have enough channels at high frequency because the spectrum of high redshift galaxies is rising with frequency. Nevertheless this spectrum has to be flatter than the spectrum of the local dust because the $300\mu$ to 1mm range is closer to the maximum of the redshifted dust emission. The statistics of the fluctuations will be an important tool which requires the coverage of a fairly large fraction of the sky. From that point of view the rise of the curves in the figures is steep meaning that the "good sky" is a large fraction of the total.

## 3 Mission concept

Measuring the cosmological background anisotropies with an accuracy better than a part in a million on scales down to ten arc minutes is a very ambitious goal. Within 10 years, balloon-borne experiments are likely to have located the first Doppler peak and measured it's amplitude with 10-30% accuracy. Their main limitation comes form the short duration of exposures available, the atmospheric effects, and difficulties in controling foregrounds contamination and systematics. On the other hand, although difficult, the goal above is achievable in the coming decade with a dedicated satellite experiment using the best detectors available today.

Indeed, the European Space Agency is studying a mission (COBRAS-SAMBA) which is aimed at the measurements described above[9]. This mission will be up for selection in the spring of 1996 to become the third medium-size mission of the Horizon 2000 plan. This concept results from the work of the ESA Science Team[10].

The payload will be passively cooled; preliminary thermal models indicate that the focal plane assembly will reach an average temperature of ∼100 K, while the telescope optical surfaces will stabilize at ∼120 K. These temperatures, in addition to the absence of the atmospheric noise and heat load due to the chosen orbit, insure that the conditions for a low and stable background are met.

---

[8] In the case of CMB fluctuations produced by topological defects like cosmic strings, the angular resolution is even more crucial, for their signature stands out all the more clearly than one goes to smaller scales.

[9] A less ambitious version is studied by the French Space Agency (CNES) in collaboration with laboratories in UK, Italy and USA. It relies only on bolometers cooled to 100 mK with the same technique as in COBRAS-SAMBA mission and plans to include only 5 wavelengths channels. It is based on a 80-cm telescope placed in a polar sun-synchronous orbit.

[10] The COBRAS/SAMBA Science Team which carried the study was : M. Bersanelli, C. Cesarsky, L. Danese, G. Efstathiou, M. Griffin, J.-M. Lamarre, M. Mandolesi, H.U. Norgaard-Nielsen, O. Pace, E. Pagana, J.-L. Puget, J. Tauber and S. Volonté. Requests for copies of the relevant report can be obtained from S. Volonté, ESA HQ, 8-10 rue Mario Nikis, PARIS CEDEX 15, FRANCE.

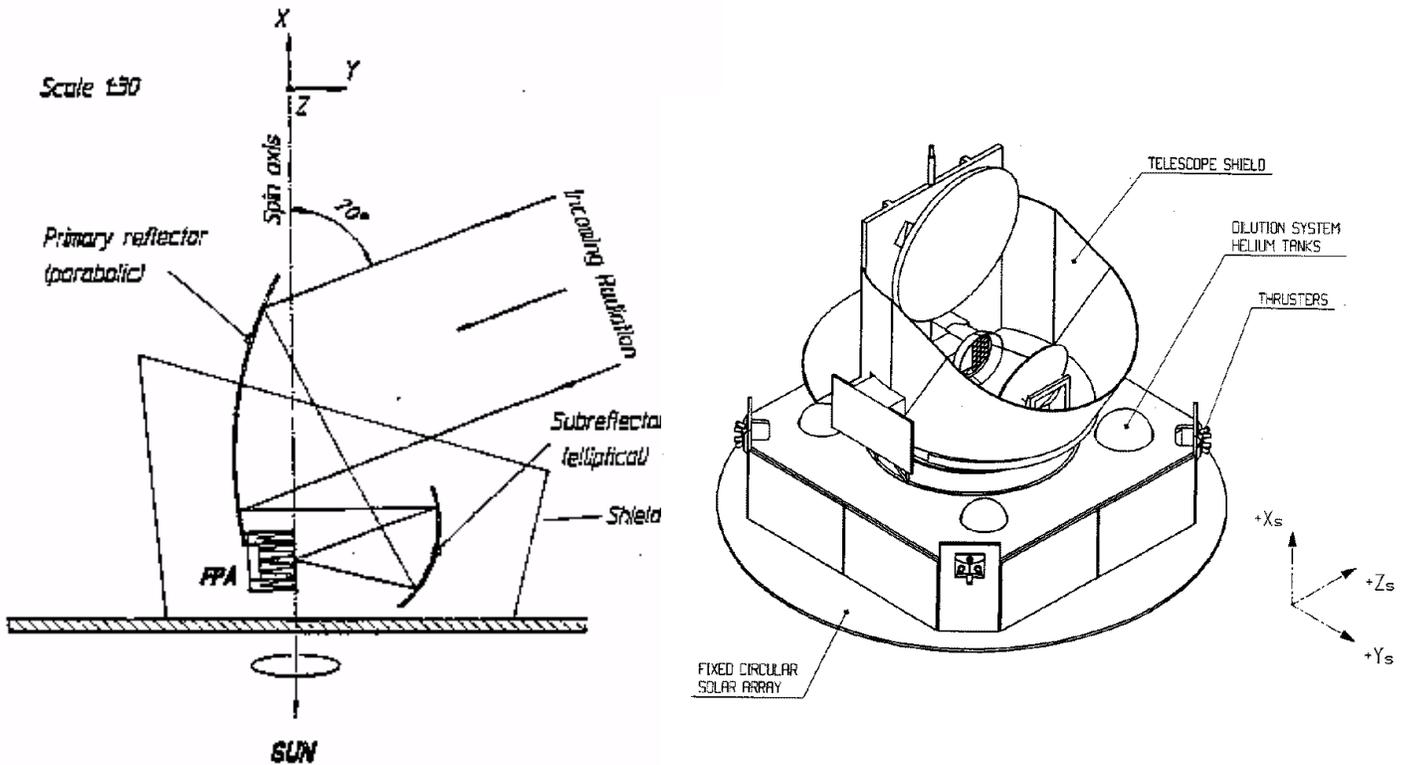

Figure 5: (a) Cross-sectional sketch of the model payload. Shown are the largest flared shield, the telescope and the focal plane assembly. (b) 3D view of the model payload.

An off-axis 1.5 meter gregorian telescope provides the necessary angular resolution and collecting area to meet the objectives of the mission. The separation of the different components of the foreground emissions requires a broad frequency coverage. The proposed mission will incorporate nine frequency bands provided by five arrays of bolometers at high frequency and four arrays of passively cooled tuned radio detectors at low frequency. The 56 bolometers require cooling to $\sim 0.1$ K, which will be achieved with an active system similar to the one foreseen for the ESA cornerstone mission FIRST. It combines active coolers reaching 4 K with a dilution refrigeration system working at zero gravity, both being developed and space-qualified in Europe. The refrigeration system will include two pressurized tanks of $^3$He and $^4$He, giving an operational lifetime of at least two years.

On the low frequency side, HEMT (High Electron Mobility Transistors) amplifiers give the required sensitivity to detect temperature anisotropies of the cosmological background at all angular scales larger than $0°.5$. The frequency and angular coverage will provide good overlap with the COBE-DMR maps.

The frequency band at which the foreground emission is the lowest ($\sim 130$ GHz) will be covered by both the tuned radio receivers and the bolometers. The fact that the tuned radio detectors are passively cooled insures that the low frequency channels can be operated for a duration limited only by spacecraft consumables.

COBRAS-SAMBA will be placed into a halo orbit around the L2 Lagrangian point of the Earth-Sun system. This location is much more favorable than a low-earth orbit, or even the L5 orbit in terms of observing efficiency, since the Earth and Moon would always be located behind the telescope, and straylight would not be a significant issue. The L2 orbit is also very favorable from the point of view of passive cooling and thermal stability.

The requirement of coverage of a large fraction of the sky is obtained by offsetting the optical axis of the telescope by 70° with respect to the spin axis of the satellite directed along the line joining the satellite and the Sun. In this nominal position, the optical axis scans a circle of diameter 60°, centered on a point in the ecliptic plane in the anti-Sun direction. In one year, the observed circle sweeps the whole ecliptic. To increase the sky coverage, the spacecraft rotation axis will be moved away from the ecliptic plane by up to ±15°.

| Center Frequency | 31.5 | 53 | 90 | 125 | 150 | 217 | 353 | 545 | 857 |
|---|---|---|---|---|---|---|---|---|---|
| Detector Type | HEMT radiometer arrays | | | | Bolometer arrays | | | | |
| Det. Temperature | ~100 K | | | | 0.1-0.15 K | | | | |
| Cooling needed | Passive | | | | Cryocooler+Dilution system | | | | |
| Number of Detectors♣ | 4 | 14 | 26 | 12 | 8 | 12 | 12 | 12 | 12 |
| Angular Resolution | 30 | 30 | 30 | 30 | 10.5 | 7.5 | 4.5 | 4.5 | 4.5 |
| Bandwidth ($\frac{\Delta\nu}{\nu}$) | 0.15 | 0.15 | 0.15 | 0.15 | 0.35 | 0.35 | 0.35 | 0.35 | 0.35 |
| $\frac{\Delta T}{T}$ Sensitivity$^\diamond$ | 6.5 | 4.0 | 3.5 | 8.5 | 2.1 | 2.2 | 13.2 | 99.3 | $> 10^3$ |
| $\frac{\Delta T}{T}$ Sensitivity$^\heartsuit$ | 2.2 | 1.3 | 1.2 | 2.8 | 0.7 | 0.7 | 4.4 | 33 | 7600 |

♣ For the HEMTS, each dual–polarization horn supports two receivers (detectors).
$\diamond$ $1\sigma$ sensitivity for a 2 years mission, over 90% of the sky.
$\heartsuit$ $1\sigma$ sensitivity for a 2 years mission, over 2% of the sky with deep integration.

Table I: Payload Characteristics. Frequencies are in GHz, angular resolutions in arc minute, $\Delta T/T$ Sensitivity in $10^{-6}$ units.

The coverage of the sky is not uniform in terms of integration time per pixel. However, the motion of the spin axis can be chosen in such a way as to give deeper integrations in chosen parts of the sky. The regions of lowest galactic emission are of particular interest for the anisotropy measurement, and many of them will be observable during the mission. Table I shows the estimated average sensitivities per pixel at the end of the two year baseline mission, for each frequency channel, assuming an observing strategy as sketched above. The pixel size for each channel is also given in Table I.

As this brief summary of the mission concept shows, a high precision determination of the small scale anisotropies of the CMB is now within reach.

# References

Andreani 1990,

Barnes, W.J., 1994, PhD thesis at the University of Chicago

Bennet, C.L. et al., 1992, *Astrophys. J.* **397**, L7

Cheng E.S. et al., 1994, *Astrophys. J.* **422**, L37

De Bernardis P. 1990,

Fischer M. et al., 1995,

Gautier, T.N., Boulanger F., Perault M., & Puget J.-L. 1992, *Astron. J.* **344**, 35

Haslam C.G.T., Quigley M.J.S., Salter C.J., 1970, *MNRAS* **147**, 405

Meinhold, P. and Lubin P.1991, *Astrophys. J.* **370**, L11

Page L.A., Cheng E.S., & Meyer, S.S., 1990, *Astrophys. J.* **355**, L1

Reach W., et al., 1995, ApJ, in press.

Reynolds, 1992, *Astrophys. J.* **392**, L35

Wright E.L. et al. 1991, *Astrophys. J.* **381**, 200